\documentclass[preprint]{elsarticle}

\usepackage{amsmath}				
\usepackage{graphicx}				
\usepackage{psfrag}				

\begin{document}
\begin{frontmatter}

\title{Dynamics of multi-frequency oscillator ensembles with resonant coupling}
\author{S. L\"uck and A. Pikovsky}
\address{
Department of Physics and Astronomy, Potsdam University, Karl-Liebknecht-Str 24-25, 14476 Potsdam, Germany}
\begin{abstract}
 We study dynamics of populations of resonantly coupled oscillators having different frequencies. Starting from the coupled van der Pol equations we derive the Kuramoto-type phase model  for the situation, where the natural frequencies of two interacting subpopulations are in relation 2:1.  Depending on the parameter of coupling, ensembles can demonstrate fully synchronous clusters, partial synchrony (only one subpopulation synchronizes), or asynchrony in both subpopulations. Theoretical description of the dynamics based on the Watanabe-Strogatz approach is developed. 
\end{abstract}


\end{frontmatter}

\section{Introduction}
Models of coupled autonomous oscillators are used to describe synchronization phenomena~\cite{Pikovsky-Rosenblum-Kurths-01,Strogatz-00} appearing in many physical~\cite{Kiss-Zhai-Hudson-02a,Wiesenfeld-Swift-95,Glova-03} 
and biological \cite{Golomb-Hansel-Mato-01,Neda-Ravasz-Brechet-Vicsek-Barabasi-00,%
Strogatz_et_al-05,Eckhardt_et_al-07} systems. In the case of a weak 
coupling, a phase model description is appropriate, leading to the famous 
Kuramoto model~\cite{Kuramoto-75,Kuramoto-84} and its modifications~\cite{Daido-92a,Sakaguchi-Kuramoto-86}. 
One of the main assumption behind the derivation of these models is that 
the oscillators are in resonance, i.e. their frequencies are close to 
each other (even when a bimodal distribution of frequencies is  considered (see \cite{Martens_etal-09,Pazo-Montbrio-09} and references there),  
one assumes that the distance between the peaks is small).  Recently, 
we considered ensembles of oscillators consisting 
of \textit{non-resonantly} coupled groups~\cite{Komarov-Pikovsky-11}, 
i.e. those with frequencies that are far from each other and far 
from resonances. 

In this paper we study synchronization effects in ensembles where different groups of oscillators are in a non-trivial resonance relation $2:1$. First, we derive general equations describing these interacting 
subpopulations in the phase approximation. Then we demonstrate 
numerically regimes of complete and partial synchrony (in the latter 
case one subpopulation synchronized while the other not). Furthermore, 
we develop a theory based on the Watanabe-Strogatz approach~\cite{Watanabe-Strogatz-93,Watanabe-Strogatz-94,Pikovsky-Rosenblum-08,Pikovsky-Rosenblum-09,Pikovsky-Rosenblum-11} that allows one
a description in terms of dynamical equations for the order parameters.

\section{Basic Model}
While typically one considers ensembles of oscillators with close frequencies that interact resonantly~\cite{Kuramoto-84},
here we focus on the two populations having natural frequencies $\omega$ and $\Omega=2\omega$.
Therefore, we spend more space than usual describing the derivation of the phase model. We start with two
coupled van der Pol oscillators 
\begin{equation}
\begin{aligned}
  \ddot{x}-\mu_1(1-x^2)\dot{x}+\omega^2x&=f_1(x,\dot x,y,\dot y)\;, \\
  \ddot{y}-\mu_2(1-y^2)\dot{y}+\Omega^2y&=f_2(x, \dot x,y,\dot y) \;, 
 \end{aligned}
 \label{eq:van_der_pol}
\end{equation}
and following a standard procedure write the averaged equations for the slow 
varying complex amplitudes $A(t)= e^{-i\omega t}(x-i\dot x/\omega)$, $B(t)= e^{-i\Omega t}(x-i\dot x/\Omega)$:
\begin{align*}
  \dot{A}&=\frac{\mu_1}{2}(1-|A|^2/4)A-\frac{i}{\omega}\langle f_1(x,\dot{x},y,\dot{y})e^{-i\omega t})\rangle\;,\\
  \dot{B}&=\frac{\mu_2}{2}(1-|B|^2/4)B-\frac{i}{\Omega}\langle f_2(x,\dot{x},y,\dot{y})e^{-i\Omega t})\rangle\;.
 \end{align*}
The  interacting terms that survive the averaging are 
those with dependence $f_1\sim e^{i\omega t}$ and
$f_2\sim e^{i\Omega t}$. Thus, due to the resonance condition $2\omega=\Omega$, $f_1$ should contain a product of $x$ and $y$, while
$f_2$ should contain the square of $x$. Therefore the simplest polynomial 
terms that yield a coupling between two oscillators are 
$f_1=c_1 xy+c_2\dot{x}y+c_3 x\dot{y}+c_4\dot{x}\dot{y}$ and $f_2=d_1 x^2+d_2\dot{x}x+d_3 \dot{x}^2$. Correspondingly,
the averaged equations can be written as
\begin{align*}
  \dot{A}&=\frac{\mu_1}{2}(1-|A|^2/4)A+\sigma_1e^{i\alpha_1} A^* B\;,\\
  \dot{B}&=\frac{\mu_2}{2}(1-|B|^2/4)B+\sigma_2e^{i\alpha_2} A^2\;,
 \end{align*}
with some complex coupling constants $\sigma_{1,2}e^{i\alpha_{1,2}}$ that can be expressed in terms of constants
$c_{1-4},d_{1-3}$. As the next step, we use smallness of $\sigma_{1,2}$ compared to $\mu_{1,2}$, 
so that the deviations of the amplitudes $|A|,|B|$ from the limit cycle values $|A|=|B|=2$ are small. Then, substituting $A=2e^{i\phi(t)}$ and
$B=2e^{i\psi(t)}$, we obtain for the phase dynamics
 \begin{align*}
   \dot{\phi}&=2\sigma_1\sin(\psi-2\phi+\alpha_1)\;,\\
   \dot{\psi}&=2\sigma_2\sin(2\phi-\psi+\alpha_2)\;.
  \end{align*}
By shifting one of the phases $\psi=\psi'-\alpha_1$ and rescaling the time variable $2(\sigma_1+\sigma_2)t= t'$
we can reduce the dynamics to a system with two parameters $\mu=\sigma_1/(\sigma_1+\sigma_2)$ and $\gamma=\alpha_1+\alpha_2$ only (we use the same letters for new variables)
 \begin{align*}
   \frac{d\phi}{dt}&=\mu\sin(\psi-2\phi)\;, \\
   \frac{d\psi}{dt}&=(1-\mu)\sin(2\phi-\psi+\gamma).
  \end{align*}
These equations describe two coupled 
phase oscillators $\phi$ and $\psi$ with a relative 
coupling strength $\mu\in[0,1]$. Because of symmetry $\phi,\psi,\gamma\to-\phi,-\psi,-\gamma$ we vary the phase shift
in the range  $\gamma\in[0,\pi]$. 
The parameters $\mu$ and $\gamma$ are functions 
of the coupling terms in Eqs.~(\ref{eq:van_der_pol}).

Now we generalize to ensembles of oscillators, assuming that each unit in a population with single frequency
$\omega$ interacts with every unit in a population with double frequency $\Omega=2\omega$: 
\begin{equation}
  \begin{aligned}
   \frac{d \phi_k}{dt}&=\frac{\mu}{N_{\Omega}}\sum^{N_{\Omega}}_{j=1}\sin(\psi_j-2\phi_k)\;, \\
   \frac{d \psi_k}{dt}&=\frac{1-\mu}{N_{\omega}}\sum^{N_{\omega}}_{j=1}\sin(2\phi_j-\psi_k+\gamma)\;. 
  \end{aligned} 
  \label{eq:ensemble}
  \end{equation}
  here $N_\omega$, $N_\Omega$ are the sizes of subpopulations.
Equations (\ref{eq:ensemble}) describe the basic model 
that we will investigate in the following. It consists of two groups of oscillators with a frequency ratio 2:1. 
Each group is composed of identical oscillators. One oscillator of a 
group is coupled to all oscillators of the other group, and vice verse. We assume that there is no interaction within one group.

\section{Dynamical regimes and their characterization}
We first present numerical results of simulations of ensemble (\ref{eq:ensemble}), setting the relative coupling strength $\mu=0.5$ and the number of oscillators 
to be equal in each group $N_{\Omega}=N_{\omega}=N$. Our main attention here is to the dependence of the dynamics
on the phase shift $\gamma$ and on different initial conditions.

\begin{figure}
 \centering
 \psfrag{xlabel2}[c][c]{time}
 \psfrag{10pi}[c][c]{$10\pi$}
 \psfrag{ylabel2}[c][c]{$\phi_k(t),\psi_k(t)$}
 \psfrag{xlabel0}[c][c]{time}
 \psfrag{ylabel0}[c][c]{$Y_j(t)$}
 \psfrag{ylabel1}[c][c]{$Z_j(t)$}
(a)\includegraphics[width=0.45\columnwidth]{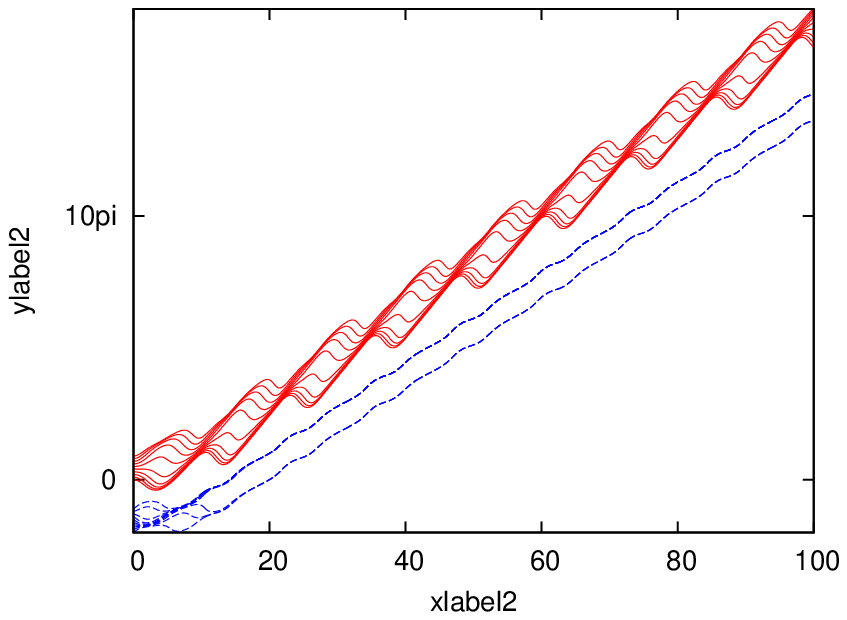} \hfill
(b)\includegraphics[width=0.45\columnwidth]{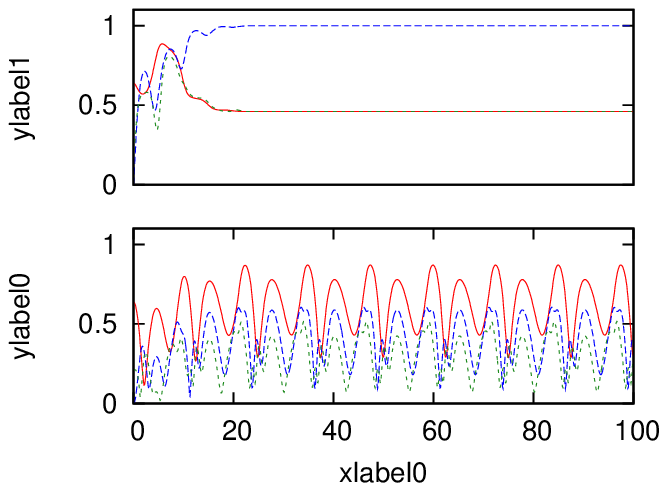}  
 \caption{(a) Phases of coupled ensembles (\ref{eq:ensemble}) 
as functions of time for  $\gamma=2.8$, $N=200$, 
and $\mu=0.5$ and the corresponding evolution of the generalized 
order parameter (b). In (a) red full lines are $\phi_k$ (only 10 
phases from the population are shown for clarity) and blue 
dashed lines are $\psi_k$. In (b) red full lines, blue dashed lines 
and green dotted lines correspond to $j=1,2,3$. }
 \label{fig:orderparam_single}
\end{figure}

We illustrate a nontrivial regime of the interaction of two populations in
Fig.~\ref{fig:orderparam_single}. 
Here, for $N=200$ and $\gamma=2.8$, by integrating Eqs.~(\ref{eq:ensemble}) we observe that  single-frequency oscillators ($\phi_k$, blue dashed curves) form two clusters that differ by $\pi$, while
double-frequency  oscillators ($\psi_k$, red full curves) remain distributed in some range of phases.

To characterize the synchronization properties and the clustering, we adopt
the Daido generalized order 
parameters~\cite{Daido-92a,Daido-93a,Daido-96}, calculated separately for double- and single-frequency ensembles:
\begin{equation} Z_j(t)=\frac{1}{N_{\omega}}\sum^{N_{\omega}}_{k=1}
 \exp\left[ij\phi_k(t)\right]\;,
 \qquad Y_j(t)=\frac{1}{N_{\Omega}}\sum^{N_{\Omega}}_{k=1}
 \exp\left[ij\psi_k(t)\right]\;.
\label{eq:orderparam}
\end{equation} 
The physical meaning of these quantities is clear from considering the case of large ensembles, then it follows from (\ref{eq:orderparam}) that $Z_j,Y_j$ are the $j$-th Fourier modes of the distributions of the phases.
While the usual Kuramoto order parameters $Z_1,Y_1$
are suitable for characterization of distributions having a single maximum (single clusters), the second order parameters
$Z_2,Y_2$ allow us to reveal 2-cluster states (distributions with two humps with the phase difference $\pi$) -- at these states these parameters have absolute value one, while the Kuramoto parameters $Z_1,Y_1$ vanish. Similarly,  $Z_3,Y_3$ are suitable for 
revealing an 
existence of three clusters with phase shift $\frac{2\pi}{3}$ (or, more generally, three-hump distributions).

In Fig.~(\ref{fig:orderparam_single}(b) the evolution
of the order parameters for the dynamics depicted in Fig.~\ref{fig:orderparam_single}(a) is shown. 
After a transient time interval, one can see that $|Z_2|=1$ while $|Z_1|<1$ and $|Z_3|<1$, as expected. We note here that in the case $|Z_2|=1$, the exact values of order parameters $|Z_{1,3}|$ are irrelevant as they 
characterize the partition between two clusters seen in Fig.~\ref{fig:orderparam_single}(a). As these clusters differ by $\pi$, this
partition  has no effect on the dynamics, where only the values $2\phi_k$ entry (cf. Eq.~(\ref{eq:ensemble})). The variations of the order parameters $|Y_j|,\;j=1,2,3$ characterize the oscillating distribution of the phases of the double-frequency oscillators.

Next, we want to characterize the dynamics of the order parameters in dependence on the parameter $\gamma$. Therefore, after a certain transient time we calculated the minimal and maximal values of the amplitudes of the 
order parameters in course of their evolution.   Hence, 
a finite interval between the maximum and the minimum characterizes a range of variations of the 
order parameter: e.g., for the regime presented in Fig.~\ref{fig:orderparam_single}, variations of  $|Y_j|$ are finite while there is no variations in 
$|Z_j|$.

\begin{figure}
 \centering
 \psfrag{xlabel0}[c][c]{$\gamma$}
 \psfrag{ylabel0}[c][c]{$Y_1$}
 \psfrag{ylabel1}[c][c]{$Y_2$}
 \psfrag{ylabel2}[c][c]{$Y_3$}
 \psfrag{ylabel3}[c][c]{$Z_1$}
 \psfrag{ylabel4}[c][c]{$Z_2$}
 \psfrag{ylabel5}[c][c]{$Z_3$}
 \psfrag{pi}[c][c]{$\pi$}
  \psfrag{pi2}[c][c]{$\pi/2$}
  \includegraphics[width=0.45\textwidth]{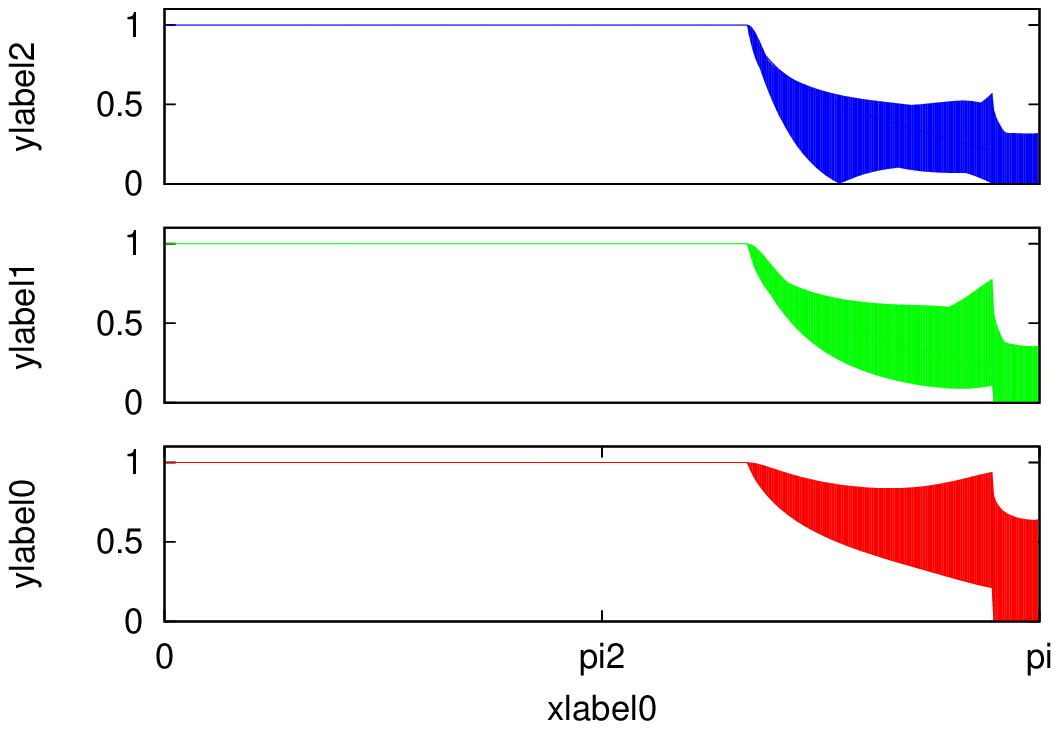}\hfill
  \includegraphics[width=0.45\textwidth]{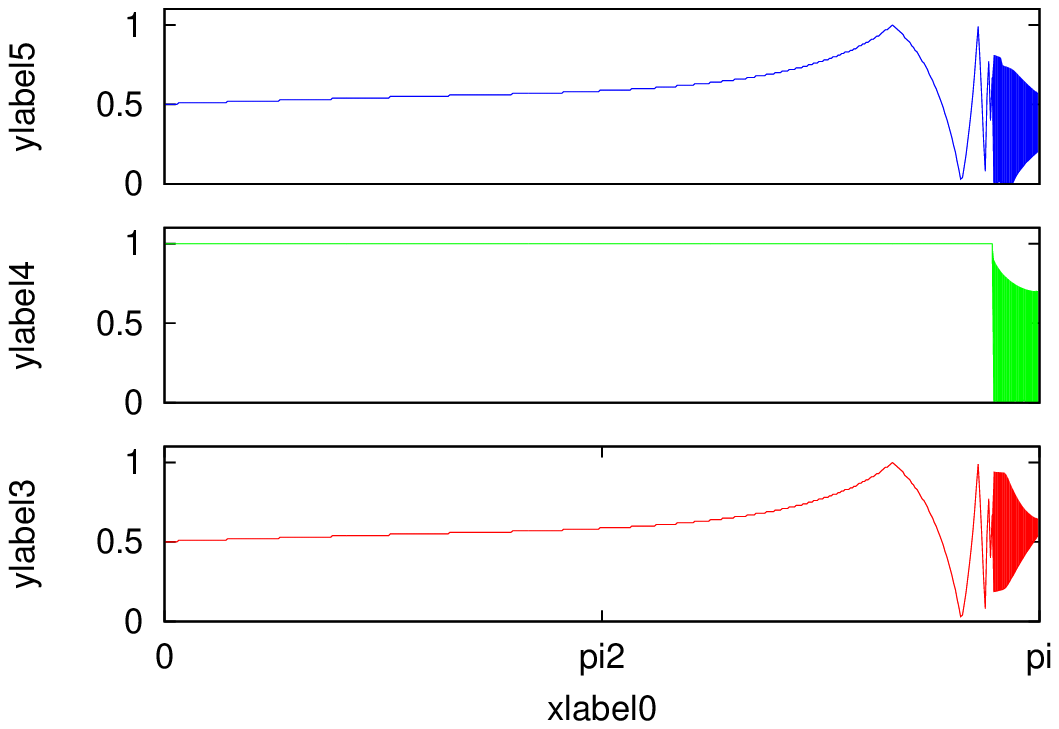}
 \caption{Variations of the order parameter $|Z_j(\gamma)|$, $|Y_j(\gamma)|$ for $N_\Omega=N_\omega=200$, $\mu=0.5$ in dependence on $\gamma$.  Here the initial distribution is a uniform one over half of the phase circle: $0<\phi_k,\psi_k<\pi$.}
 \label{fig:equal_dist}
\end{figure}

\begin{figure}
 \centering
\psfrag{xlabel0}[c][c]{$\gamma$}
 \psfrag{ylabel0}[c][c]{$Y_1$}
 \psfrag{ylabel1}[c][c]{$Y_2$}
 \psfrag{ylabel2}[c][c]{$Y_3$}
 \psfrag{ylabel3}[c][c]{$Z_1$}
 \psfrag{ylabel4}[c][c]{$Z_2$}
 \psfrag{ylabel5}[c][c]{$Z_3$}
 \psfrag{pi}[c][c]{$\pi$}
  \psfrag{pi2}[c][c]{$\pi/2$}
   \includegraphics[width=0.45\textwidth]{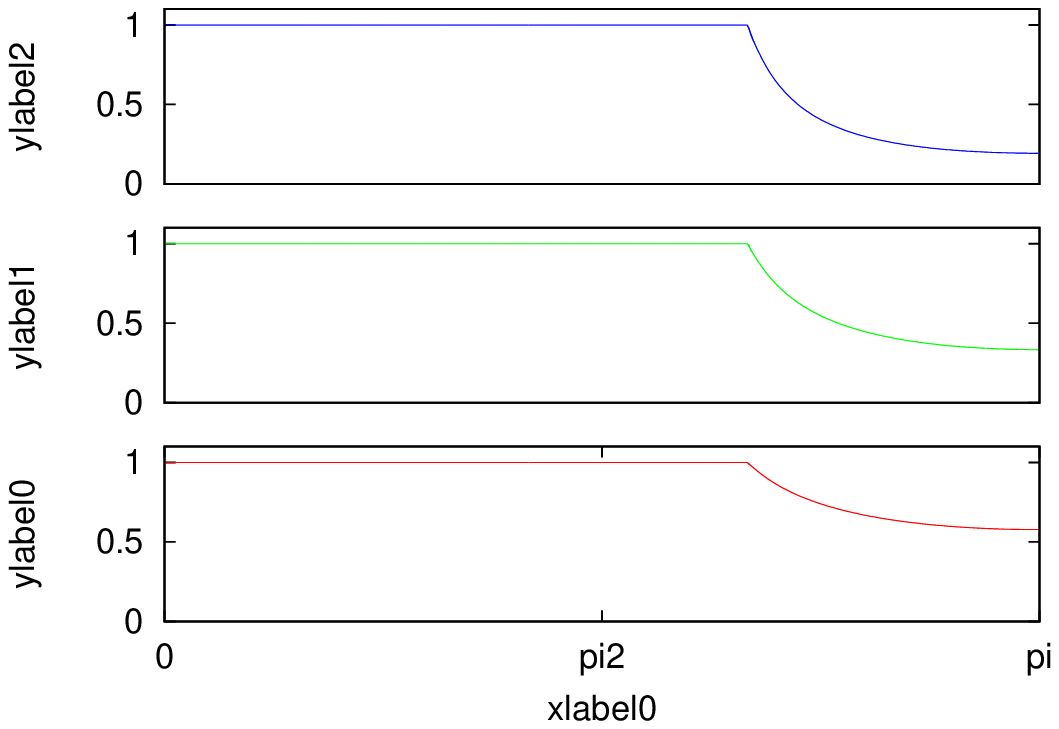}\hfill
  \includegraphics[width=0.45\textwidth]{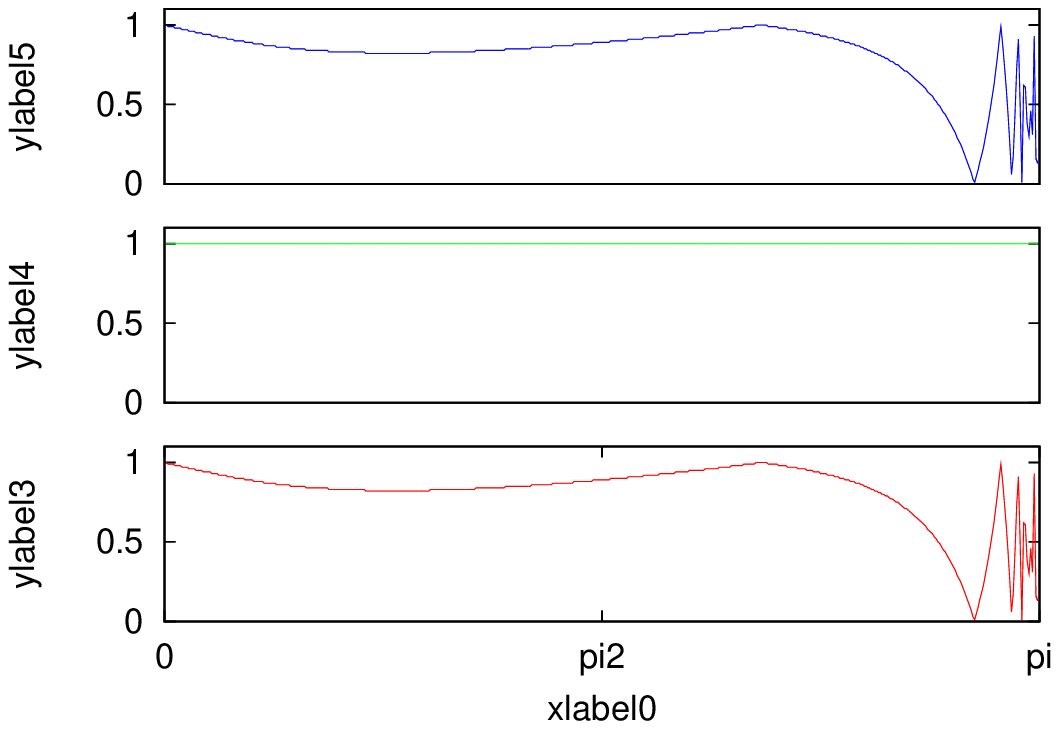}
 \caption{The same as in fig. (\ref{fig:equal_dist}), but with a different initial condition: The oscillators were initially transformed according to the WS theory (see Eq.~\ref{eq:wstr}) below).}
 \label{fig:ws-dist}
\end{figure}

We have found that dynamical regimes strongly depend on initial conditions.
In Figs.~\ref{fig:equal_dist},\ref{fig:ws-dist} we show the variations of the order parameters for two sets of initial conditions: in Fig.~\ref{fig:equal_dist} both phases are uniformly distributed in the interval $[0,\pi)$; in another set Fig.~\ref{fig:ws-dist} the initial conditions are chosen specially, according to the theory we develop below in Section~\ref{sec:ws_system}. As discussed above, the most relevant are the data for $Y_1$ and $Z_2$, while values of $Z_1,Z_3$ giving a partition between subclusters of single-phase oscillators are irrelevant for the dynamics. 
For $0<\gamma<\gamma_{c1}\approx 2.094$ both these order parameters are one, what means that in both subpopulations full synchrony establishes. For $\gamma_{c1}<\gamma<\pi$ the order parameter $|Y_1|<1$ what means asynchrony in the double-frequency subpopulation. For initial conditions in Fig.~\ref{fig:ws-dist} the single-frequency subpopulation is synchronized in this range, while for the setup of Fig.~\ref{fig:equal_dist} the single-frequency population is synchronous up to $\gamma_{c2}\approx 2.98$ and asynchronous for $\gamma>\gamma_{c2}$. Thus, the coupled ensembles demonstrate regimes of full synchrony for $\gamma<\gamma_{c1}$, partial synchrony for $\gamma_{c1}<\gamma<\gamma_{c2}$ and asynchrony (unless special initial conditions are chosen) for $\gamma_{c2}<\gamma<\pi$.

\begin{figure}
 \centering
 \psfrag{xlabel0}[c][c]{$\text{Re}(Y_1,Z_2)$}
\psfrag{ylabel0}[c][c]{$\text{Im}(Y_1,Z_2)$}
 \psfrag{xlabel1}[c][c]{$\text{Re}(Y_1)$}
\psfrag{ylabel1}[c][c]{$\text{Im}(Y_1)$}
 \psfrag{xlabel2}[c][c]{$\text{Re}(Z_2)$}
\psfrag{ylabel2}[c][c]{$\text{Im}(Z_2)$}
   (a)\includegraphics[width=0.3\textwidth]{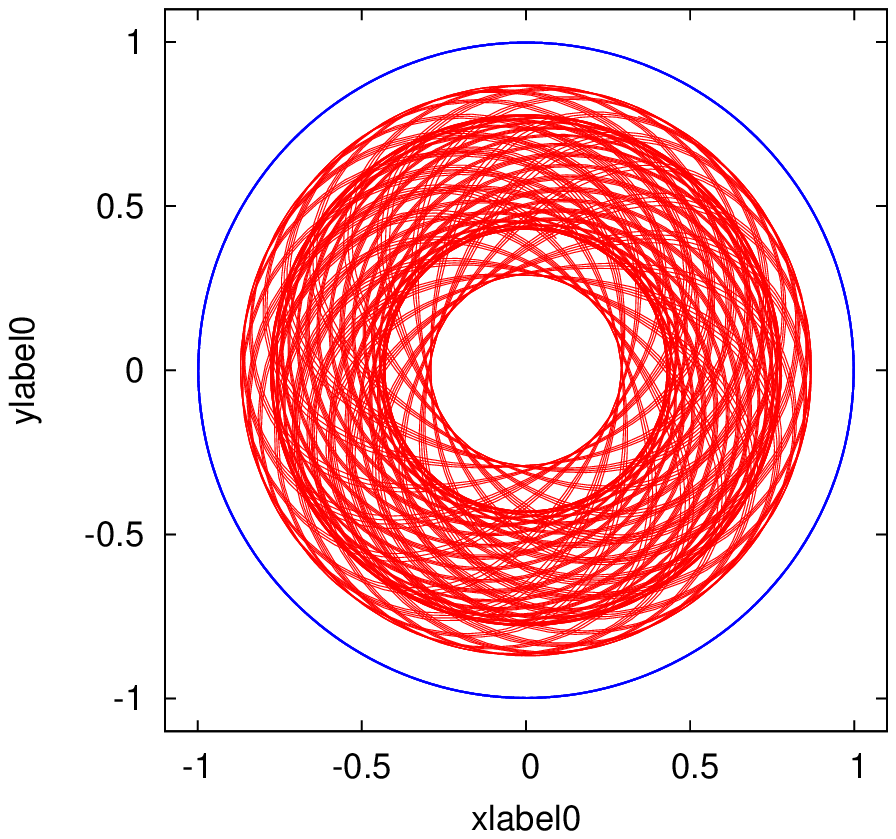}\hfill
   (b)\includegraphics[width=0.28\textwidth]{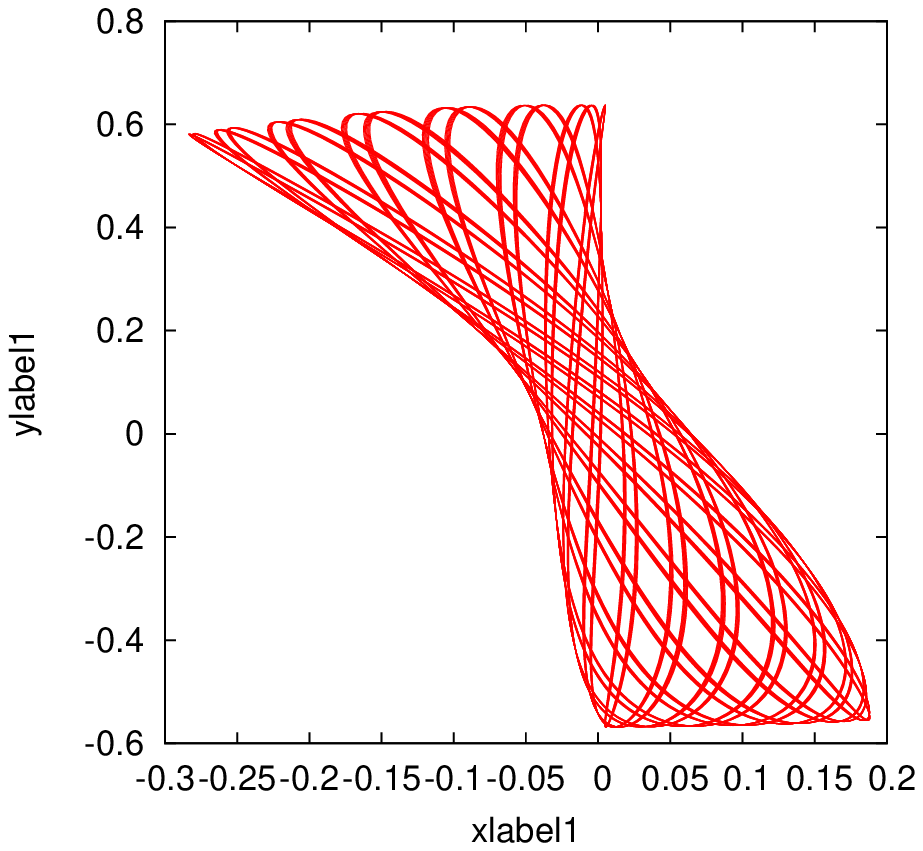}\hfill
   (c)\includegraphics[width=0.28\textwidth]{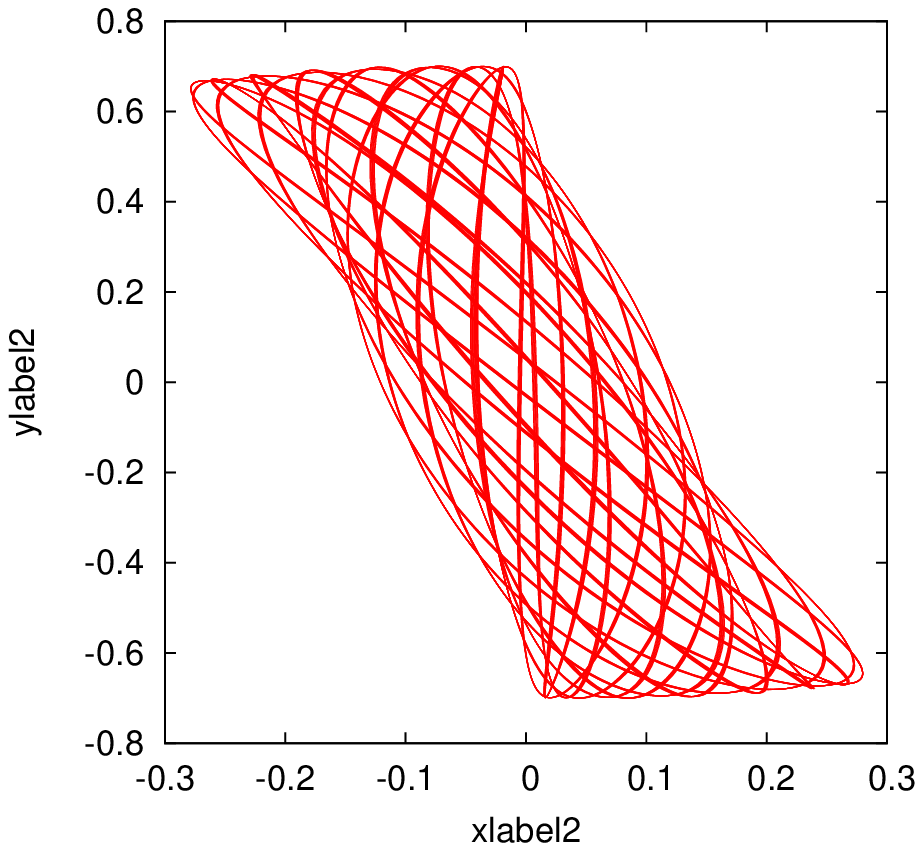}
 \caption{Projections of the dynamical regimes on the planes of order parameters, for $N=200$ and $\mu=0.5$. (a): $\gamma=2.8$, here $|Z_2|=1$ (blue circle) while $Y_1$ varies in some range (red curve). (b),(c): For $\gamma=3.08$ both order parameters vary quasiperiodically.}
 \label{fig:qp}
\end{figure}

To reveal the type of dynamics in regimes of partial synchrony and asynchrony, we show in Fig.~\ref{fig:qp} the two-dimensional projections of trajectories on the planes of main order parameters. One can see that in both cases the dynamics is two-frequency quasiperiodic; this is also confirmed by calculations of Poincar\'e maps, on which the attractor form one-dimensional lines.

\section{Watanabe-Strogatz ansatz} 
 \label{sec:ws_system}
In this section we apply the WS ansatz allowing us to describe the system of coupled phase oscillators with a few global variables. In this way one is able to analyze such systems analytically. We begin with a sketch of the WS theory
according to~\cite{Pikovsky-Rosenblum-11}, for an original formulation see~\cite{Watanabe-Strogatz-93,Watanabe-Strogatz-94}.  One starts with
an ensemble of $N$ identical phase oscillators with frequencies $\omega(t)$ driven by a force $H(t)$ according to
\[
 \dot{\varphi_k}=\omega(t) +\mbox{Im}[H(t)e^{-i\varphi_k}]\;,\qquad	k=1,...,N>3\;,
\]
and performs a transformation to new microscopic variables $\vartheta_k$ and global variables $z,\zeta$ ($z$ is complex, $\zeta$ is real) according to
\begin{equation}
   e^{i\varphi_k}=\frac{z+e^{i(\vartheta_k+\zeta)}}{z^* e^{i(\vartheta_k+\zeta)}+1}
   \label{eq:wstr}
   \end{equation}
with additional conditions $\sum e^{ i\vartheta_k}=0$ and $\text{Re}( \sum e^{ i2\vartheta_k})=0$. Then, the new microscopic phases $\vartheta_k$ are constants of motions provided the macroscopic variables $z,\zeta$ satisfy the WS equations\begin{align*}
  \frac{dz}{dt} &= i\omega(t) + \frac{1}{2} \left( H(t) - z^2H^*(t)  \right)\;,\\
  \frac{d\zeta}{dt} &= \omega(t) + \mbox{Im}(z^*H(t))\;.
\end{align*}
As discussed in~\cite{Pikovsky-Rosenblum-11}, the complex variable $z$ is roughly proportional (but not exactly equal) to the Kuramoto order parameter of the population, while $\zeta$ is a shift between individual phases and the phase of $z$. Indeed, from the transformation (\ref{eq:wstr}) it follows that 
\begin{equation}
\langle e^{i\phi}\rangle=
\frac{1}{N}\sum_{k=1}^{N}
\frac{z+e^{i(\vartheta_k+\zeta)}}{1+z^*e^{i(\vartheta_k+\zeta)}}=\langle e^{i\phi}\rangle(z,\zeta,\vartheta_k)\;.
\label{eq:wsop}
\end{equation}
Only in the case of a uniform distribution of constants $\vartheta_k$ on the interval $[0,2\pi)$ and in the thermodynamic limit $N\to\infty$, the variable $z$ coincides with the Kuramoto order parameter, because the dependence on $\zeta$ and $\vartheta_k$ in (\ref{eq:wsop}) disappears and $z=\langle e^{i\varphi}\rangle$.

To apply the WS theory to our system (\ref{eq:ensemble}) 
 we perform a transformation $2\phi_k=\theta_k$ and rewrite the system as
\begin{equation}
\begin{aligned}
  \dot{\theta_k} &= 2\mu \langle\sin\left(\psi_j-\theta_k\right)\rangle \qquad \quad\;\;\; = 2 \mu \mbox{Im}(Y_1 e^{-i\theta_k}) \\
  \dot{\psi_k} &= (1-\mu) \langle\sin\left(\theta_j-\psi_k+\gamma\right)\rangle\vspace*{-3mm} = (1-\mu) \mbox{Im}(Z_2 e^{-i\psi_k} e^{i\gamma} )
\end{aligned}
\label{eq:ens2}
\end{equation}
Now we are able to apply the WS transform with two sets of WS variables $z_{1,2}$, $\zeta_{1,2}$, for subpopulations of oscillations having single and double frequencies, correspondingly. As discussed above, the consideration extremely simplifies in the case of a uniform distribution of the corresponding constants of motion $\vartheta_k$
and in the thermodynamic limit, where a simple representation of the order parameters via the WS variables holds:
\begin{equation}
 z_1=Z_2 \quad \mbox{and} \quad z_2=Y_1\;.
 \label{eq:mf-ws}
\end{equation}
In this case, the closed WS system reads
 \begin{equation}
  \begin{aligned}
   \frac{dz_1}{dt} &= \mu\left( z_2-z_1^2 z_2^* \right)\;,\\ 
   \frac{dz_2}{dt} &= \frac{(1-\mu)}{2}\left( z_1e^{i\gamma}-z_2^2 z_1^*e^{-i\gamma} \right)\;.
  \end{aligned}
\label{eq:wssys}
 \end{equation}
With introduction of amplitudes $\rho_{1,2}=|z_{1,2}|$ and the phase difference $\Psi=\text{arg}(z_2)-\text{arg}(z_1)$ we obtain a three-dimensional system
\begin{equation}
\begin{aligned}
\dot{\rho_1} &= \mu \left(1-\rho_1^2\right)\rho_2\cos\left(\Psi\right)\;, \\
 \dot{\rho_2} &= \frac{1}{2}\left( 1-\mu\right)\rho_1\left( 1-\rho_2^2 \right)\cos\left( \Psi-\gamma\right) \;,\\
 \dot{\Psi} &= \frac{\mu-1}{2}\rho_1\frac{1+\rho_2^2}{\rho_2}\sin\left( \Psi-\gamma\right)-\mu\frac{1+\rho_1^2}{\rho_1}\rho_2\sin\Psi \;.
 \end{aligned}
\label{eq:WS-system}
\end{equation}

We first discuss the steady states in system (\ref{eq:WS-system}) and their stability. 
The steady state corresponding to a full synchrony is 
\begin{equation}
\rho_1^{(0)}=\rho_2^{(0)}=1,\quad\tan\Psi^{(0)}=
\frac{(\mu-1)\sin\gamma}{(\mu-1)\cos\gamma-2\mu}\;.
\label{eq:fss}
\end{equation}
It is stable if $\cos\gamma>\text{max}(\frac{\mu-1}{2\mu},\frac{2\mu}{\mu-1})$.
(There is another state with $\Psi=\Psi^{(0)}+\pi$ which is unstable).

 Two stable steady states with partial synchrony are possible. For parameter values
$-1<\cos\gamma<\frac{\mu-1}{2\mu}; \mu>1/3$ the state 
\[
\rho_1^{(1)}=1,\quad \rho_2^{(1)}=\sqrt{\frac{\mu-1}{4\mu\cos\gamma+1-\mu}},\quad\Psi^{(1)}=-\pi/2+\gamma
\]
 is stable while for  $-1<\cos\gamma<\frac{2\mu}{\mu-1}; \mu<1/3$ another state
\[
\rho_2^{(2)}=1,\quad \rho_1^{(2)}=\sqrt{\frac{\mu}{(\mu-1)\cos\gamma-\mu}},\quad\Psi^{(2)}=\pi/2
\]
 is stable. The fully asynchronous state $\rho_1=\rho_2=0$ is unstable for $\gamma<\pi$.

The special case $\gamma=\pi$ deserves a separate analysis. In this case the dynamics is described by equations
\begin{align}
 \dot{\rho_1} &= \mu \left(1-\rho_1^2\right)\rho_2\cos\Psi\;, \label{eq:ham1}
\\
 \dot{\rho_2} &= \frac{1}{2}\left( 1-\mu\right)\rho_1\left( 1-\rho_2^2 \right)\cos\Psi\;,\label{eq:ham2}
\\
 \dot{\Psi} &= \left(\frac{1-\mu}{2}\rho_1\frac{1+\rho_2^2}{\rho_2}
 -\mu\frac{1+\rho_1^2}{\rho_1}\rho_2\right)\sin\Psi \;.
\label{eq:ham3}
 \end{align}
One can see that this system has two integrals. One is obtained by dividing (\ref{eq:ham1}) and (\ref{eq:ham2}) and integration; another one is 
obtained by first using the first integral to express $\rho_2(\rho_1)$,  and then dividing (\ref{eq:ham1}) and (\ref{eq:ham3}) whith subsequent integration.
Thus, the resulting dynamics is conservative and periodic.

The analytical results above explain Fig.~\ref{fig:ws-dist}, where the initial conditions have been chosen according to Eq.~(\ref{eq:wstr}) with a uniform distribution of the constants $\vartheta_k$. In Fig.~\ref{fig:ws-dist} one observes a regime of full synchrony corresponding to steady state $\rho_{1,2}^{(0)},\Psi^{(0)}$ for $\gamma<\gamma_{c1}=2\pi/3$ and the steady state 
$\rho_{1,2}^{(1)},\Psi^{(1)}$ for $\gamma>2\pi/3$, in according with relations above and $\mu=1/2$. For $\gamma=\pi$ one observes oscillations of the order parameters.

In order to explain Fig.~\ref{fig:equal_dist} we have to go beyond the assumption used at the derivation of (\ref{eq:wssys}), namely of the uniform distribution of constants $\vartheta_k$ and of thermodynamic limit. In general case, instead of (\ref{eq:mf-ws}) we have to use (\ref{eq:wsop}) for $Z_2=Z_2(z_1,\zeta_1,\vartheta_k^{(1)})$ and $Y_1=Y_1(z_2,\zeta_2,\vartheta_k^{(2)})$.
Now the full system of equations is
\begin{equation}
\begin{aligned}
\dot z_{1} &= \mu \left( Y_1 - z_1^2Y_1^*  \right)\\
\dot\zeta_1 &= 2\mu\mbox{Im}(z_1^*Y_1(t))\\
\dot z_{2} &= \frac{1-\mu}{2} \left( Z_2 - z_2^2Z_2^*  \right)\\
\dot\zeta_2 &= (1-\mu) \mbox{Im}(z_2^*Z_2)
\end{aligned}
\label{eq:wsfull}
\end{equation} 
The fully synchronous state where both populations are completely synchronized is the same as in system (\ref{eq:WS-system}), because as one can see from (\ref{eq:wsop}), for $|z|=1$ we again have $\langle e^{i\phi}\rangle=z$, but all other states are generally different. In particular, partially synchronous regimes are not steady states but quasiperiodic ones as in Figs.~\ref{fig:orderparam_single},\ref{fig:equal_dist},\ref{fig:qp}. We discuss their stability in the next section.

\section{Stability of partially synchronous state}
\begin{figure}
\centering    
\psfrag{xlabel0}[c][c]{$\gamma$}
\psfrag{ylabel0}[c][c]{$\lambda_{ev}$}
\includegraphics[width=0.4\textwidth]{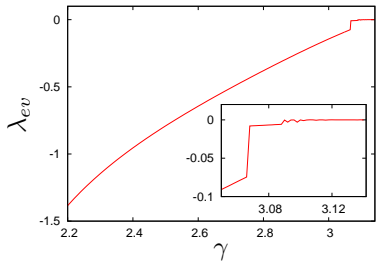}
 \caption{Evaporation exponent calculated for $N_\omega=200$, $\mu=0.5$, in dependence on
 $\gamma$. Initial conditions: phases $\phi_k$ uniformly distributed in $[0,\pi)$. The inset shows region around $\gamma=\pi$. For $\gamma<3.05$ the exponent is large and negative, while for $\gamma>3.1$ the exponent vanishes.}
 \label{fig:evap_exponent}
\end{figure}

The WS theory above allows us to describe analytically the transition full synchrony $\to$ partial synchrony as loss of stability of the fully synchronous state (\ref{eq:fss}), but the analysis of the transition from partial synchrony (where one ensemble forms a synchronous cluster while other one is asynchronous)
is more involved as it deals with quasiperiodic states.
Therefore we apply here a direct numerical method for determining stability of clusters -- calculation of so-called evaporation Lyapunov exponents~\cite{Kaneko-94,Pikovsky-Popovych-Maistrenko-01}. We assume, according to numerics, that the single-frequency oscillators are different, while the double-frequency oscillators form a synchronous cluster, i.e. in  Eq.~(\ref{eq:ensemble}) $\psi_1=\psi_2=\ldots=\psi_{N_\Omega}=\tilde\psi$. Then, in the limit $N_{\Omega}\to\infty$, the deviation of one element from the cluster is governed by
\[
\frac{d}{dt} \delta\psi=\delta\psi\frac{\partial}{\partial \tilde\psi}
\frac{1-\mu}{N_{\omega}}\sum^{N_{\omega}}_{j=1}\sin(2\phi_j-\tilde\psi+\gamma)\;.
\]
Thus, the growth rate of $\delta\psi$ is determined by the evaporation exponent
\[
\lambda_{ev}=\left\langle\frac{\partial}{\partial \tilde\psi}
\frac{1-\mu}{N_{\omega}}\sum^{N_{\omega}}_{j=1}\sin(2\phi_j-\tilde\psi+\gamma)
\right\rangle\;.
\]
The results of calculation of this exponent are presented in Fig.~\ref{fig:evap_exponent}. One can see that the synchronized cluster of double-frequency oscillators is strongly stable for $\gamma<3.1$ while for $\gamma>3.1$ the evaporation exponent vanishes. This means marginal stability of the synchronized state, what is consistent with the observation of non-synchronous dynamics for appropriate initial conditions. 

\section{Conclusion}

In this paper we have studied a novel model of \textit{resonantly interacting multi-frequency oscillator populations}. As the simplest setup we have chosen a situation where oscillators are divided in two subpopulations: some have natural frequency $\omega$ while other ones have the double frequency $\Omega=2\omega$. Such a setup can be easily generalized to a general case of two subpopulations in a resonance $\Omega:\omega=m:n$. Moreover, one could study many 
subpopulations having resonantly related frequencies $\omega_1:\omega_2:\omega_3:\cdots=m_1:m_2:m_3:\cdots$, the generalization of equations (\ref{eq:ensemble}) to this case is 
straightforward.

Our main finding is that depending on the parameters of the coupling of two 
ensembles, one observes regimes of full synchrony (both subpopulations form fully synchronous clusters), partial synchrony (one subpopulation synchronized while other is asynchronous) and no synchrony (both subpopulations asynchronous). The latter two regimes demonstrate quasiperiodic dynamics. To analyse these regimes we applied the Watanabe-Strogatz theory and derived the equations for macroscopic variables describing distributions of oscillators in subpopulations. This allowed us to identify the transition from full to partial synchrony as a transcritical bifurcation of this system. To characterize the transition from partial synchrony to asynchrony we used the method of evaporation Lyapunov exponents.

In this paper we restricted our attention to the case of identical oscillators in subpopulations. For the study of non-identical ensembles the powerful Ott-Antonsen theory~\cite{Ott-Antonsen-08,Ott-Antonsen-09} can be adopted, these results will be reported elsewhere.

\bibliographystyle{elsarticle-num}

\end{document}